\documentclass[12pt]{iopart}

\usepackage{color}
\usepackage{graphicx}
\usepackage{xspace}

\newcommand{\TK}{$T_{\rm K}$\xspace}
\newcommand{\TN}{$T_{\rm N}$\xspace}
\newcommand{\TC}{$T_{\rm C}$\xspace}
\newcommand{\EF}{$E_{\rm F}$\xspace}
\newcommand{\CeAgGe}{CeAg$_2$Ge$_2$\xspace}
\newcommand{\CeRhGe}{CeRh$_2$Ge$_2$\xspace}
\newcommand{\CeRuGe}{CeRu$_2$Ge$_2$\xspace}
\newcommand{\CeCuGe}{CeCu$_2$Ge$_2$\xspace}
\newcommand{\CeNiGe}{CeNi$_2$Ge$_2$\xspace}
\newcommand{\CeCuSi}{CeCu$_2$Si$_2$\xspace}
\newcommand{\YbCoSi}{YbCo$_2$Si$_2$\xspace}
\newcommand{\YbRhSi}{YbRh$_2$Si$_2$\xspace}
\newcommand{\YbCoRhSi}{Yb(Co$_{x}$Rh$_{1-x}$)$_2$Si$_2$\xspace}
\newcommand{\YbIrSi}{YbIr$_2$Si$_2$\xspace}
\newcommand{\cf}{$c$-$f$\xspace}
\newcommand{\OC}{$\sigma(\omega)$\xspace}
\newcommand{\R}{$R(\omega)$\xspace}
\newcommand{\DCOG}{$\Delta \langle \omega \rangle$\xspace}
\newcommand{\DSW}{$\Delta SW$\xspace}
\newcommand{\Tep}{$T_{el\mathchar`-ph}$\xspace}

\begin{document}

\title{
Optical evidence of local and itinerant states in Ce- and Yb-heavy-fermion compounds
}

\author{
Shin-ichi Kimura$^{1,2,3}$, 
Yong Seung Kwon$^{4,1}$, 
Cornelius Krellner$^{5}$,
and
J\"org Sichelschmidt$^{6,1}$ 
}

\address{$^1$ Graduate School of Frontier Biosciences, Osaka University, Suita, Osaka 565-0871, Japan}
\address{$^2$ Department of Physics, Graduate School of Science, Osaka University, Toyonaka, Osaka 560-0043, Japan}
\address{$^3$ Department of Electronic Structure, Institute for Molecular Science, Okazaki, Aichi 444-8585, Japan}
\address{$^4$ Department of Emerging Materials Science, DGIST, Daegu 711-873, Republic of Korea}
\address{$^5$ Kristall- und Materiallabor, Physikalisches Institut, Goethe-Universit\"at Frankfurt, 
Max-von-Laue Stra\ss e 1, D-60438, Frankfurt am Main, Germany}
\address{$^6$ Max Planck Institute for Chemical Physics of Solids, Nothnitzer Stra\ss e 40, 01187 Dresden, Germany}
\ead{kimura@fbs.osaka-u.ac.jp}

\vspace{10pt}
\begin{indented}
\item[]
Version \today
\end{indented}

\begin{abstract}

The electronic properties of Cerium (Ce) and ytterbium (Yb) intermetallic compounds may display a more local or more itinerant character depending on the interplay of the exchange interactions among the $4f$ electrons and the Kondo coupling between $4f$ and conduction electrons.
For the more itinerant case, the materials form heavy-fermions once the Kondo effect is developed at low temperatures.
Hence, a temperature variation occurs in the electronic structure that can be traced by investigating the optical conductivity (\OC) spectra.
Remarkably, the temperature variation in the \OC spectrum is still present in the more localized case, even though the Kondo effect is strongly suppressed. 
Here, we clarify the local and itinerant character in the electronic structure by investigating the temperature dependence in the \OC spectra 
of various Ce and Yb compounds with a tetragonal ThCr$_2$Si$_2$-type crystal structure.
We explain the temperature change in a unified manner.
Above temperatures of about 100~K, the temperature dependence of the \OC spectra is mainly due to the electron-phonon interaction, 
while the temperature dependence below is due to the Kondo effect.
\end{abstract}

%
%
%
%
%

%
\section{Introduction}

Intermetallic compounds with cerium (Ce) and ytterbium (Yb) ions are known as heavy-fermion (HF) materials, 
in which the effective carrier mass ($m^*$) increases up to thousand times heavier than that of free carriers 
with decreasing temperature owing to the Kondo effect,
which develops below a characteristic temperature called Kondo temperature \TK~\cite{Hewson1993}.
Since the effective carrier mass can be described as $m^*=kdk/dE(k)|_{k=k_{\rm F}}$ of the dispersion curve of the conduction band $E(k)$, where $k$ is the wavenumber, the band dispersion is expected to be modified with temperature.
Whereas the change of the density of states of the conduction band is contained in the optical conductivity (\OC) and scanning tunneling spectroscopy measurements~\cite{Kirchner2020},
the band dispersions $E(k)$ can be directly detected by angle-resolved photoelectron spectroscopy (ARPES),
which is a method now being widely used in HF materials~\cite{Im2008, Koitzsch2008, Okane2009, Vyalikh2010, Nakatani2018}.
However, since ARPES is very sensitive to solid surfaces due to low-energy electrons' detection, the results are sometimes inconsistent with bulk properties such as transport.
An \OC spectrum is a photon-in and photon-out measurement and thus reflects the band shape due to direct transitions with the momentum transfer $q=0$ using low-energy photons in the IR and THz regions~\cite{Degiorgi1999}.
Although the $4f$ states and conduction bands of Ce and Yb compounds are renormalized by many-body effects, the \OC peaks in the middle-infrared region (``mid-IR peak'') are consistent with the unoccupied (occupied) $4f$ peak positions in Ce (Yb) compounds derived from density functional theory (DFT) band calculations with a self-energy shift from the Fermi level (\EF)~\cite{Kimura2009-1,Kimura2009-2}.
This result suggests that DFT band calculation is a good approximation for the \OC spectrum, unlike for ARPES of HF systems, which is explained by the band calculations including many-body effects within the dynamical mean-field theory (DFT+DMFT)~\cite{Chul2013, Zhang2016, Jang2020}.
The reason for this is that photoemission spectroscopy, including ARPES, observes the state with one electron extracted from the ground state, whereas \OC spectra show the ground state with a matched number of electrons, {\it i.e.}, the charge is conserved in the material, which is presumably due to the small many-body effect after photo-excitation.

\begin{figure*}[t]
\begin{center}
\includegraphics[width=0.9\textwidth]{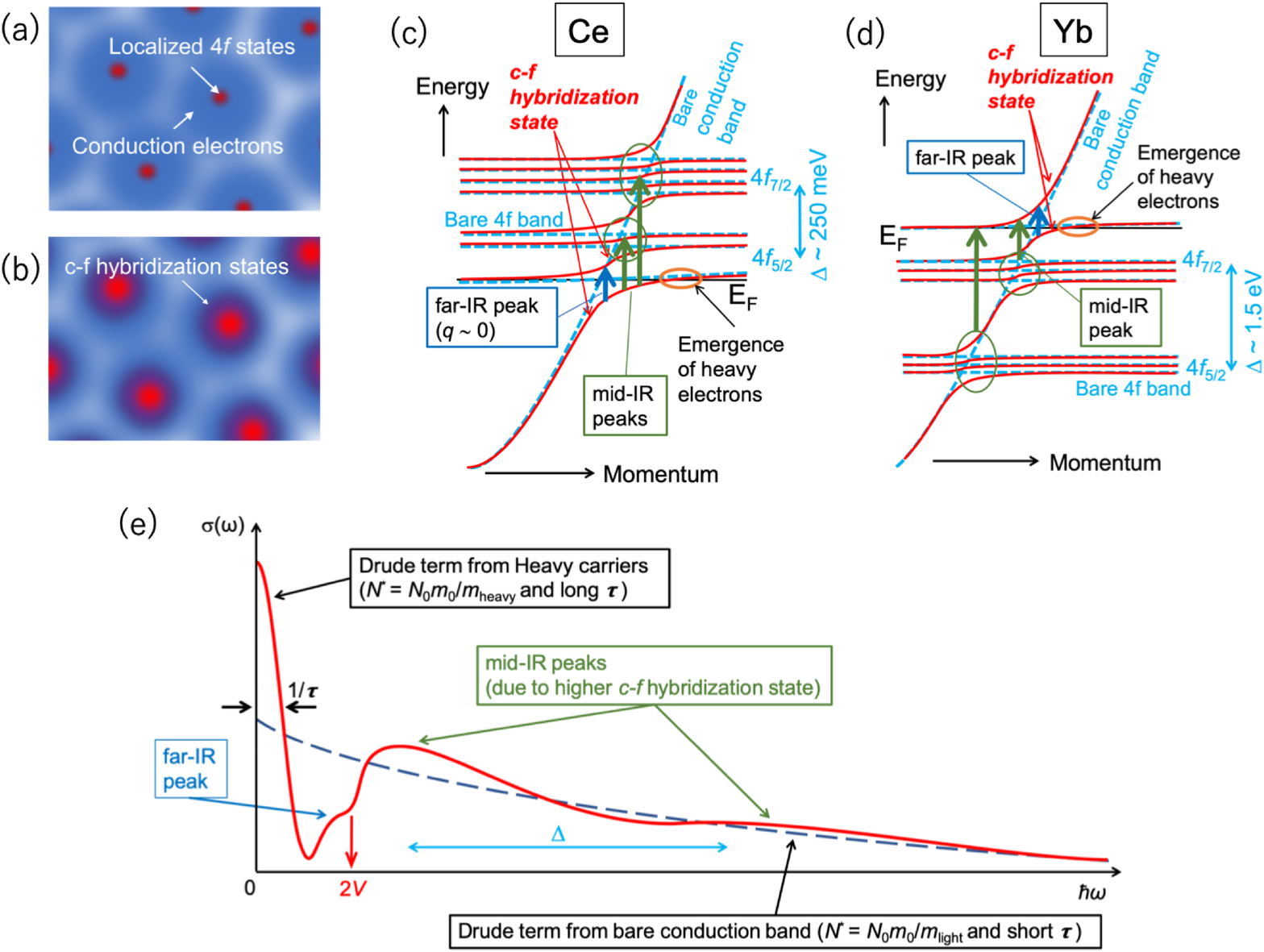}
\end{center}
\caption{
(a, b) Schematic spatial image of localized (a) and itinerant (b) $4f$ states in Ce and Yb intermetallic compounds.
In the localized $4f$ states (a), the overlap between $4f$ states and conduction electrons is small, but in the itinerant case (b), the \cf hybridization is developed, and the overlap becomes large.
(c, d) Schematic band structure of Ce (c) and Yb (d) compounds.
The dashed and solid lines indicate the localized and itinerant band structure, respectively.
The \cf hybridization state appears in the itinerant band structure.
The spin-orbit splitting energies $\Delta$ in Ce and Yb compounds are about 250~meV and 1.5~eV, respectively.
The split band at the $4f$ level is a simple example of hybridization, and in reality, 
the hybridization intensity depends on the symmetry of the bands. 
Vertical arrows indicate possible optical transitions.
(e) Schematic optical conductivity (\OC) spectra from localized band structure (a dashed line) and itinerant \cf hybridized one (a solid line).
The horizontal scale is linear.
The difference in the spectrum between Ce and Yb compounds appears in the double peak with the splitting energy of 250~meV in Ce compounds and single peak in Yb compounds owing to the larger spin-orbit splitting than the IR energy region of below 1~eV. 
}
\label{fig:SchematicFigure}
\end{figure*}

Figure~1 illustrates how the \OC spectroscopy can be useful in the study of HF systems for characterizing the bulk ground state~\cite{Kimura2011-1,Kimura2011-2,Kimura2011-3,Iizuka2012,Kimura2000,Okamura2005,Kimura2013,Kimura2016-1}.
A schematic image of localized and itinerant $4f$ electrons in real space is shown in Figs.~\ref{fig:SchematicFigure}(a,b), 
and a schematic diagram of the band structure of Ce and Yb compounds is shown in Figs.~\ref{fig:SchematicFigure}(c,d).
The localized $4f$ electrons are bound at the atomic position in the real space 
with a small spatial overlap with the spatially extended conduction ($c$) electrons (Fig.~\ref{fig:SchematicFigure}(a)).
On the other hand, in the momentum space, the localized $4f$ electrons have a narrow energy width 
and are spread over the whole momentum space, while the $c$ band is hardly hybridized with the $4f$ electrons (Figs.~\ref{fig:SchematicFigure}(c,d)).
In Ce compounds, the Ce$^{3+}$~$4f^{1}$ is stable in the localized state, and the $4f$ state and the $c$ band are hardly hybridized (dashed line in Fig.~\ref{fig:SchematicFigure}(c)),
whereas in Yb compounds, the Yb$^{3+}$~$4f^{13}$ ($4f^1_h$: $4f$ hole) state is localized (Fig.~\ref{fig:SchematicFigure}(d)).
As a result, the overlap of the wave functions of the $4f$ and $c$ states becomes small, and the optical transition probabilities of 
$|4f^n c^m \rangle + \hbar\omega \to |4f^{n+1} c^{m-1} \rangle$ for Ce compounds or 
$|4f^n c^m \rangle + \hbar\omega \to |4f^{n-1} c^{m+1} \rangle$ for Yb compounds are weak.
Now, when the $4f$ electrons are hybridized with the conduction electrons (\cf hybridization) (Fig.~\ref{fig:SchematicFigure}(b)), 
the hybridization band expands spatially.
In other words, the overlap between the wave functions of $4f$ electrons and conduction electrons becomes large.
The optical transition probabilities of $|4f^n c^m \rangle + \hbar\omega \to |4f^{n\pm1} c^{m\mp1} \rangle$ 
(actually, the optical transitions between the bonding and antibonding orbitals of the \cf hybridization band) become larger.

In heavy-fermion systems, $4f$ electrons have a localized character at temperatures higher than \TK, but, with decreasing temperature, 
some of the $4f$ states are hybridized with the $c$ band (\cf hybridization) and behave itinerantly (solid line in Fig.~\ref{fig:SchematicFigure}(c)).
Then, the curvature of the bare $c$-band at \EF becomes flat at low temperature 
due to the \cf hybridization band's appearance, resulting in heavier quasiparticles.
Besides, since the \cf hybridization band has an energy gap of twice the hybridization strength ($2V$) on the \EF, 
where $V$ is the hybridization energy,
a shoulder structure of several meV to several ten~meV appears, corresponding to the size of the \cf hybridization gap.
Furthermore, in HF systems, a characteristic peak called mid-IR peak appears.
The origin of the mid-IR peak is also considered to be due to the \cf hybridization.
A schematic diagram of the change in the \OC spectrum due to \cf hybridization is shown in Fig.~\ref{fig:SchematicFigure}(e).
In the localized state, at high temperature, a Drude structure from the bare $c$-band appears (dashed line, without the \cf hybridization). 
As the temperature decreases, the simple form changes to a complex shape with a renormalized Drude peak, 
a far-IR peak due to the transition across the \cf hybridization gap, and a mid-IR peak structure due to the development of the \cf hybridization.
The mid-IR peak is assigned to the optical transition of $|4f^n c^m \rangle + \hbar\omega \to |4f^{n+1} c^{m-1} \rangle$ for Ce compounds 
and $|4f^n c^m \rangle + \hbar\omega \to |4f^{n-1} c^{m+1} \rangle$ for Yb compounds.
The spin-orbit splitting of the $4f$ states of Ce and Yb ions must appear in the \OC spectra.
The spin-orbit splitting in Ce compounds is about 250~meV, which corresponds to the peak spacing of the double-peak structure in the mid-IR peak.
On the other hand, in the Yb compound, since the spin-orbit splitting energy of the $4f$ states is about 1.5~eV, only one peak appears in the mid-IR peak region, 
while the other peak is expected to appear in the visible light region.
However, it is difficult to observe it because it is overlapped with other absorption structures.

As for the mid-IR peak, Okamura {\it et al.} found that the peak energy is proportional to the effective mass of the heavy quasiparticle, 
which is another feature of heavy-fermion systems~\cite{Okamura2007}.
On the other hand, Kimura {\it et al.} showed that 
the behavior of the mid-IR peak changes could be explained by DFT calculations in the itinerant phase~\cite{Kimura2009-1,Kimura2009-2}, 
even though unoccupied states cannot be fully trusted in a DFT calculation because the unoccupied states cannot be optimized or adjusted during the iteration to self-consistency. 
Across a quantum critical point (QCP), however, the behavior changes to that is explained by using the local character with an on-site Coulomb repulsion in the $4f$ states~\cite{Kimura2016-2}.
In other words, by investigating the appearance of the mid-IR peak, we can study the localized and itinerant characters of $4f$ states, 
as well as the typical electronic states of HF systems.

\begin{table*}[t]
\label{table:sample}
\begin{center}
\caption{
Magnetic ordering temperatures (\TN, \TC) and Kondo temperature (\TK) of samples used for this paper~\cite{Endstra1993,Matsumoto2011,Gegenwart2002,Hossain2005,Klingner2011}.
}
    \begin{tabular}{ c | cccccc}
    \hline
    Sample & \CeAgGe & \CeRhGe & \CeRuGe & \CeCuGe & \CeNiGe & \CeCuSi \\ \hline
    \TN & 5--8~K & 15~K & 8.5~K & 4.1~K & -- & -- \\
    \TC &&& 8.0~K && \\
    \TK & -- & -- & -- & -- & several~K & 10~K \\ \hline
    \end{tabular}

    \begin{tabular}{ c | cccc}
    \hline
    Sample &  & \YbCoRhSi &  & \YbIrSi \\
    & $x=1$ &  $x=0.27$ & $x=0$ & \\ \hline
    \TN & 1.65~K & 1.3~K & 0.072~K & -- \\
    \TK & -- & 7~K & 25~K & 40~K \\ \hline
    \end{tabular}
  \end{center}
  \end{table*}

Many \OC studies have shown that this mid-IR peak is related to the \cf hybridization~\cite{Singley2002,Okamura2004,Mema2005,Singley2002,Chen2016,Bachar2016}.
However, the relationship between the temperature at which the mid-IR peak is formed, and the \TK has not been investigated systematically.
In this study, we investigated the relationship between the temperature dependence in the \OC spectra in the mid-IR region 
across the \TK in HF materials with the same tetragonal ThCr$_2$Si$_2$-type crystal structure of Ce compounds 
(Ce$M_2$Ge$_2$ ($M$~=~Ag, Rh, Ru, Cu, Ni), \CeCuSi) and of Yb compounds (\YbCoRhSi ($x = 0, 0.27, 1$), \YbIrSi).
The magnetic ordering and Kondo temperatures of these materials are listed in Table
~\ref{table:sample}~\cite{Endstra1993,Matsumoto2011,Gegenwart2002,Hossain2005,Klingner2011}.
These materials cover the overall physical property change from the local to itinerant characters in the Doniach phase diagram~\cite{Doniach1977}.
As a result, no mid-IR peaks were observed in the materials with strongly localized $4f$ levels (\CeAgGe and \YbCoSi~\cite{Gruner2012,Guttler2014}), 
and peaks indicating absorption into the levels shifted to the higher energy side due to the on-site Coulomb interaction $U$ acting on the $4f$ electrons were observed.
The slight temperature dependence of the \OC structure of these materials is concluded 
to originate from the thermal effect of the electron-phonon interaction.
On the other hand, for materials with strong itinerancy, the mid-IR peaks are visible even at room temperature, much higher than \TK.
Moreover, the temperature at which the mid-IR peaks begin to appear does not scale with \TK.
This result suggests that the mid-IR peaks only represent the \cf hybridization intensity due to the Kondo effect as the temperature decreases, 
but the \cf hybridization band's formation is not directly related to \TK.

\section{Method}

Polycrystalline samples of Ce$M_2$Ge$_2$ ($M =$ Ag, Rh, Ru, Cu, Ni) and single-crystalline of \YbCoRhSi ($x=0, 0.27$) were synthesized 
by tetra-arc melting and flux methods, respectively, 
and the surfaces were well-polished using 0.3~$\mu$m grain-size Al$_{2}$O$_{3}$ lapping film sheets to obtain shiny surfaces 
for the optical reflectivity (\R) measurements.
Near-normal incident \R spectra were accumulated in a wide photon-energy range of 2~meV -- 30~eV 
using a Martin-Puplett-type FTIR for the THz region ($\hbar\omega=2-30$~meV, FARIS-1, JASCO Co., Ltd.),
a Michelson-type FTIR for the IR region (20~meV~--~1.5~eV, FT/IR-6100, JASCO Co., Ltd.),
and the beamline 7B of a synchrotron radiation facility, UVSOR (1.2~--~30~eV)
to ensure an accurate Kramers-Kronig analysis (KKA)~\cite{Kimura2013}.
To obtain \OC via KKA of \R, the spectra were extrapolated using the Hagen-Rubens function below the lowest energy measured, 
and the free-electron approximation $R(\omega) \propto \omega^{-4}$ above the highest energy~\cite{DG}.
\OC spectra of Ce$M_2$Ge$_2$ at 8--10~K have already been reported~\cite{Kimura2016-2}.
In this paper, we will report the temperature-dependent \OC spectra of Ce$M_2$Ge$_2$ and \YbCoRhSi ($x=0, 0.27$) 
compared with those of CeCu$_2$Si$_2$~\cite{Sichelschmidt2013}, \YbRhSi~\cite{Kimura2006}, and \YbIrSi~\cite{Iizuka2010} 
to check the relation of the Kondo temperature to the temperature-dependent spectral change.

\section{Results}

\subsection{Ce compounds}

\begin{figure}[t]
\begin{center}
\includegraphics[width=0.8\textwidth]{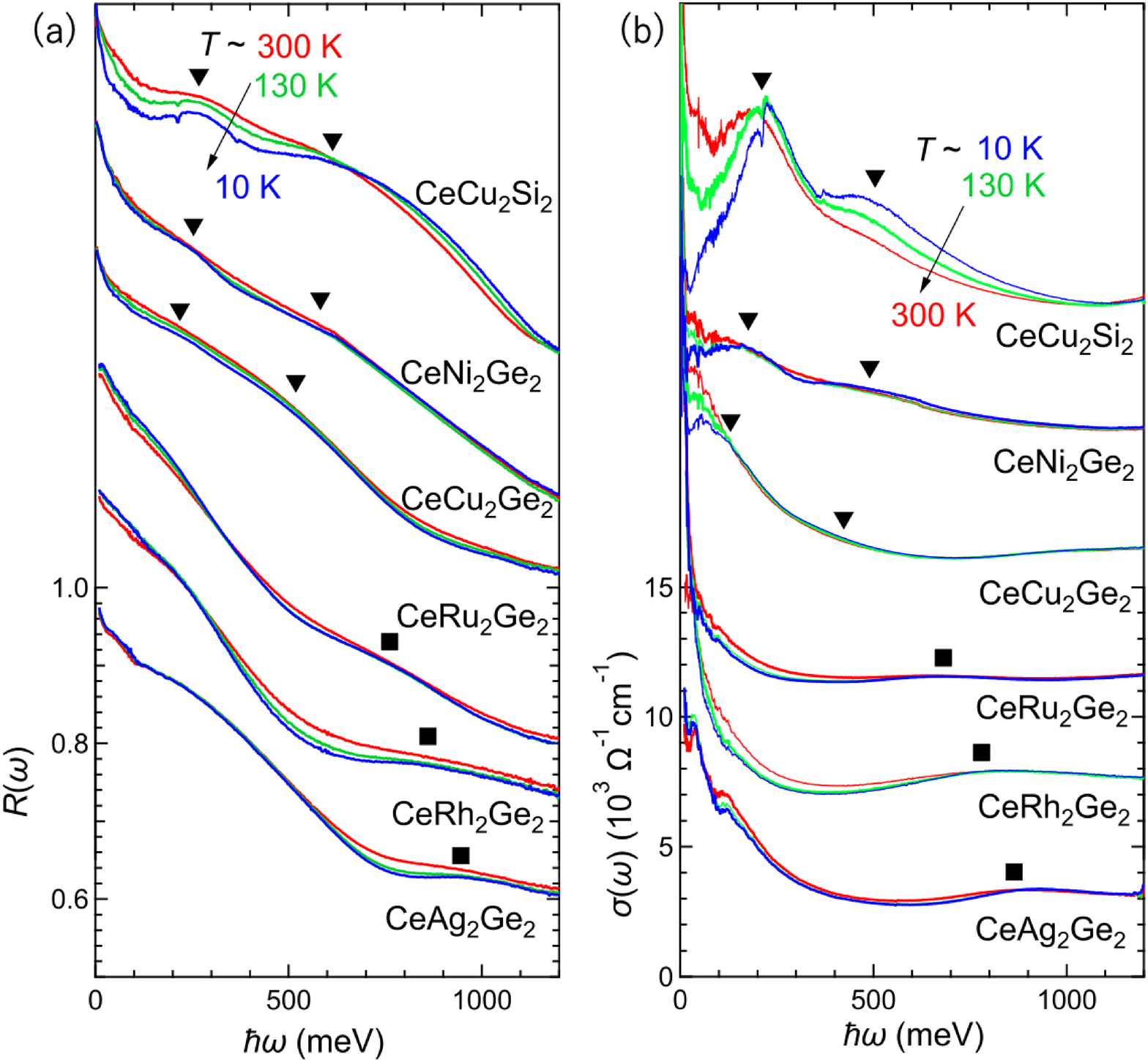}
\end{center}
\caption{
Temperature-dependent optical reflectivity (\R) spectra (a) and optical conductivity (\OC) spectra (b) of Ce$M_2$Ge$_2$ ($M =$Ag, Rh, Ru, Cu, Ni) and \CeCuSi.
The spectrum of \CeCuSi was taken from Ref.~\cite{Sichelschmidt2013}.
The baselines of each \R and \OC spectra are shifted by $0.15$ and $4.5\times10^3~\Omega^{-1}{\rm cm}^{-1}$, respectively.
Solid triangles and squares indicate the mid-IR peaks shown in Fig.~\ref{fig:SchematicFigure}(e) and the optical transitions to the unoccupied $4f$ states with the on-site Coulomb interaction ($U$-activated peak), respectively~\cite{Kimura2016-2}.
}
\label{fig:Ce}
\end{figure}

Figure~\ref{fig:Ce} shows the temperature dependence of the \R spectra and \OC spectra of the Ce compounds.
It can be seen that the \R spectra (Fig.~\ref{fig:Ce}(a)) change more or less clearly with temperature for all the samples shown in this study.
It should be noted that since it is difficult to directly compare the \R spectrum with the electronic state, we will use the \OC spectrum for the detailed discussion.
Among these, \CeCuSi, \CeNiGe with heavy electrons, and \CeCuGe with antiferromagnetic ordering and being located very close to a QCP, show a slight double-peak structure (marked by solid triangles) in the mid-IR region.
However, for the more $4f$-localized \CeRuGe, \CeRhGe, and \CeAgGe, the double-peak structure is not visible, and instead, 
a broad peak appears and shifts to higher energies as the localization increases (marked by solid squares)~\cite{Kimura2016-2}.
Clear temperature dependence is also observed in the \OC spectra of all materials (Fig.~\ref{fig:Ce}(b)).
In particular, the mid-IR double-peak intensity increases when lowering temperature and shifts to higher energy by increasing the itinerant nature in \CeCuGe, \CeNiGe, and \CeCuSi, see solid triangles.
Also, the temperature dependence of \OC increases and develops at lower temperatures.
On the other hand, the peak structure shown by solid squares, which indicates localization, appears to vary only in peak width, with little change in peak intensity.


\subsection{Yb compounds}
\begin{figure}[t]
\begin{center}
\includegraphics[width=0.8\textwidth]{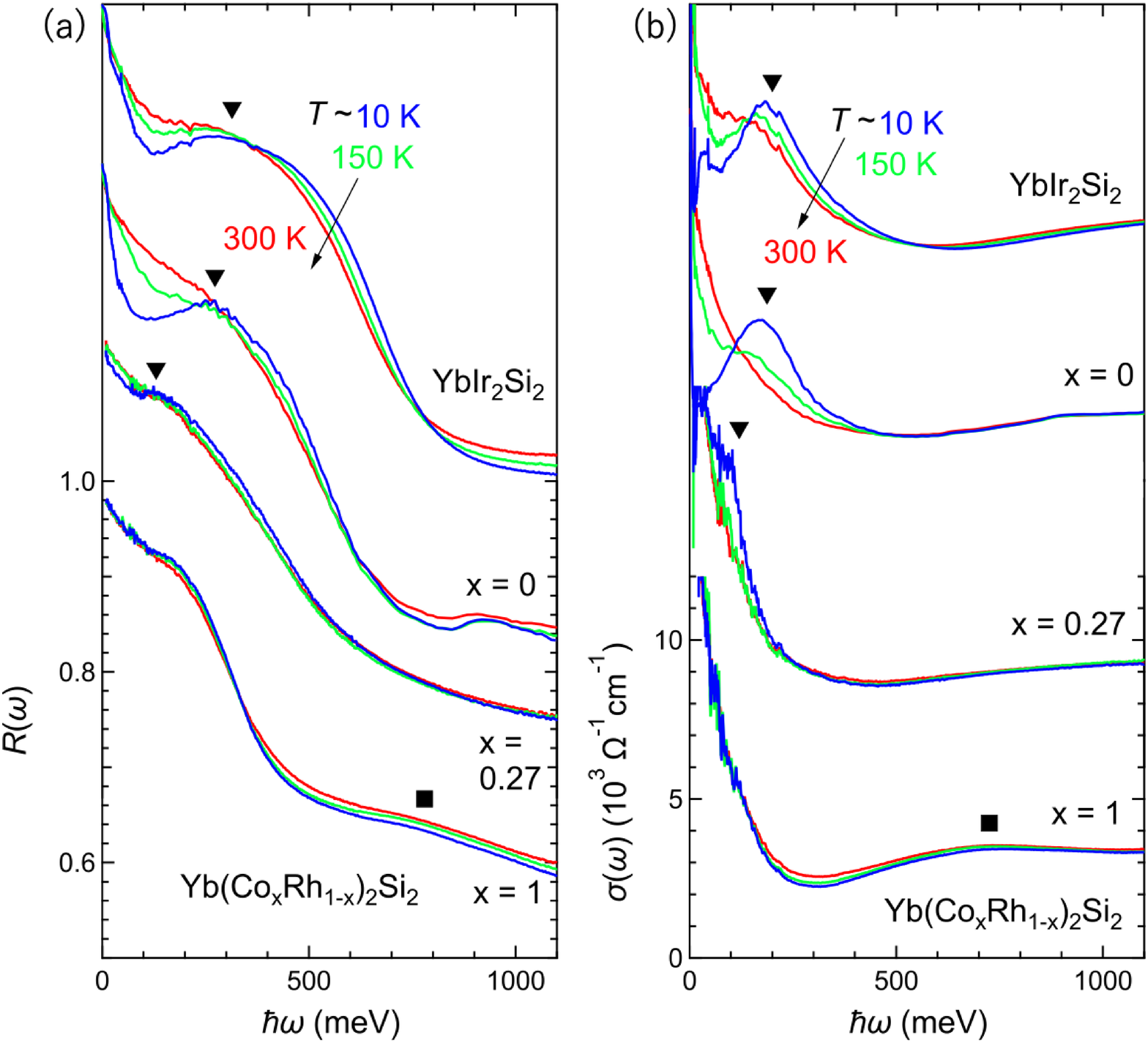}
\end{center}
\caption{
Temperature-dependent optical reflectivity (\R) spectra (a) and optical conductivity (\OC) spectra (b) of \YbCoRhSi ($x=0, 0.27, 1$) and \YbIrSi.
The spectra of \YbRhSi and \YbIrSi were taken from Ref.~\cite{Kimura2006, Iizuka2010}.
The baselines of each \R and \OC spectra are shifted by $0.17$ and $6.0\times10^3~\Omega^{-1}{\rm cm}^{-1}$, respectively.
Solid triangles and squares indicate the mid-IR peaks shown in Fig.~\ref{fig:SchematicFigure}(e) and the optical transitions from the occupied $4f$ states with the on-site Coulomb interaction ($U$-activated peak), respectively.
}
\label{fig:Yb}
\end{figure}

Figure~\ref{fig:Yb} shows the temperature dependence of the \R spectra and \OC spectra of Yb compounds.
Like Ce compounds, \YbIrSi and \YbRhSi, which are in the itinerant phase and very close to the QCP, respectively, show a clear mid-IR peak (marked by solid triangles), indicating a robust \cf hybridization.
There appears to be a small mid-IR peak in \YbCoRhSi ($x = 0.27$) located in the slightly localized region.
This result suggests that the \cf hybridization also works on the slightly localized side of the QCP.
This is consistent with the behavior in the Ce compounds shown above.
However, in the case of \YbCoSi, which is located further to the localized side, the mid-IR peak disappears.
Instead, an on-site-$U$-activated peak (marked by solid square) appears, which is also observed in localized Ce compounds,
but the temperature change in the \OC spectra becomes more pronounced.
The mid-IR peak of \YbIrSi and \YbCoRhSi ($x=0, 0.27$) (solid triangles) develops with decreasing temperature, while the on-site-$U$-activated peak of \YbCoSi (solid square) only becomes narrower with no change in peak intensity.
This result is the same as in the case of Ce compounds,
and suggests that the mid-IR peak is related to the development of the Kondo effect at low temperatures.

\section{Discussion}

\begin{figure}[t]
\begin{center}
\includegraphics[width=0.8\textwidth]{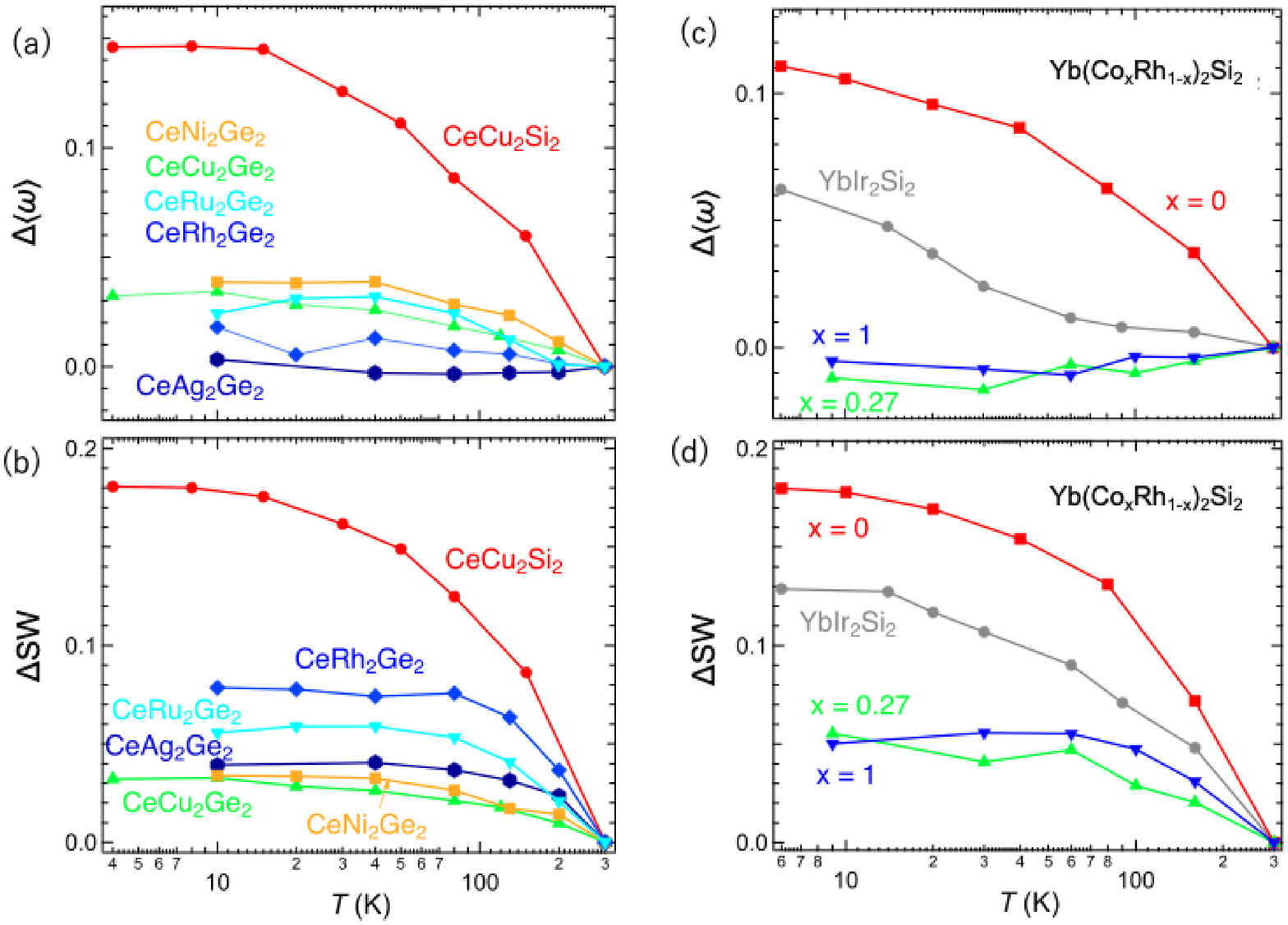}
\end{center}
\caption{
Relative temperature dependence of the change of the center of gravity (\DCOG) (a, c) and the spectral weight (\DSW) (b, d) of Ce (a, b) and Yb (c, d) compounds.
These values were normalized to their room-temperature values.
}
\label{fig:Analysis}
\end{figure}

It is commonly observed in Ce and Yb HF-compounds that the mid-IR peak (solid triangles in Figs.~\ref{fig:Ce} and \ref{fig:Yb}) due to \cf hybridization develops towards low temperatures in itinerant materials.
In contrast, the on-site-$U$-activated peak (solid squares) in localized materials does not change in intensity with lowering the temperature but instead shows a decreasing line width.
Here, we try to separate the effect of \cf hybridization and other effects from these results.
In general, the electron-phonon interaction intensity can be changed by temperature.
When the magnitude of the electron-phonon interaction changes by changing the temperature, 
the peak position in the spectrum does not change so much, but only the width does,
because the peak position and width originate from the \cf hybridization and the electron-phonon interaction, respectively.
The fact that the peak position is unchanged means that the center of gravity (\DCOG) of the spectrum does not change with temperature, and the fact that the line width changes implies that the difference of the spectral weight changes (\DSW).
It should be noted that Drude peaks also become sharp with decreasing temperature, which will cause the low-energy shift of both \DCOG and \DSW.
This effect will be visible at the lowest energy region in \OC spectra where a renormalized Drude peak appears,
so the effect is not large to \DCOG and \DSW.

To check these expectations, the change of the \DCOG and the \DSW normalized to their room-temperature values were evaluated by using the following functions:
\[
\Delta \langle \omega \rangle = \left[ \langle \omega(T) \rangle - \langle \omega({\rm 300~K}) \rangle \right] / \langle \omega({\rm 300~K}) \rangle ,
\]
where
\[
\langle \omega(T) \rangle = \int_{\omega_1}^{\omega_2} \sigma(\omega, T)\omega d\omega / \int_{\omega_1}^{\omega_2} \sigma(\omega, T) d\omega .
\]
and
\[
\Delta SW = \int_{\omega_1}^{\omega_2} |\sigma(\omega, T) - \sigma(\omega, {\rm 300~K})| / \sigma(\omega, {\rm 300~K}) d\omega .
\]
The integration range was set as $\omega_1 = 0~{\rm eV} \leq \omega \leq \omega_2 = 0.8~{\rm eV}$, 
where the spectral change in the lower energy region is almost recovered.
The obtained temperature dependence of these parameters is shown in Fig.~\ref{fig:Analysis}.

Firstly, we discuss the behaviors of the strongly localized \CeAgGe, \CeRhGe, and \YbCoSi.
According to Fig.~\ref{fig:Analysis}(a,c), the \DCOG of these materials is almost $0$ or negative at all temperatures.
This result suggests that the electron-phonon interaction is the primary origin of the temperature variation.
The negative \DCOG suggests that the Drude peak becomes narrow with decreasing temperature, indicating increasing 
relaxation time~\cite{DG}.
On the other hand, for \DSW, shown in Fig.~\ref{fig:Analysis}(b,d), there is a temperature dependence at $T~\geq~100~{\rm K}$ (=~\Tep), and it is almost constant at $T~\leq$~\Tep.
This temperature of 100~K is not the Debye temperature~\cite{Hartmann2009}, 
however, the observation of a similar temperature dependence in the La- or Lu-reference materials~\cite{Kimura1994} 
clarifies that this is not related to $4f$ magnetism.
Therefore, the origin of these materials' temperature variation does not originate from the Kondo effect but mainly from the electron-phonon interaction.

On the other hand, \CeCuSi, \CeNiGe, \YbIrSi, and \YbRhSi are in the itinerant regime or the localized regime close to the QCP, 
both \DCOG and \DSW increase with decreasing temperature even below \Tep.
This result implyies that the \cf hybridization develops at low temperatures.
It should be noted that above \Tep, the effect of the electron-phonon interaction should also be taken into account, 
and the effect cannot be separated from the \cf hybridization effect.
In these itinerant materials, the temperature dependences of both \DCOG and \DSW start at or higher than room temperature, 
which seems to be independent of the \TK.
A similar thermal effect in the \OC spectra has been reported from not only other \OC spectra~\cite{Mema2005,Singley2002,Bachar2016} but also ARPES results of \CeNiGe~\cite{Nakatani2018}, \YbRhSi~\cite{Kummer2015}, and CeCoIn$_5$~\cite{Jang2020}.
Therefore, the electronic structure change due to the Kondo effect is considered to start at a temperature much higher than \TK. 

In \CeRuGe, located at the middle of the above materials, \DCOG increases toward lower temperatures.
This phenomenon represents a \cf hybridization development under discussion above, while the \DSW remains nearly constant below \Tep.
This result is probably due to the weak \cf hybridization strength.
A mid-IR double-peak structure is not visible even at low temperature, but an on-site-$U$-activated peak is visible (Fig.~\ref{fig:Ce}, solid squares).

On the other hand, for \YbCoRhSi ($x=0.27$), the \DCOG is almost constant (rather decreasing) with decreasing temperature, 
while the \DSW is observed to be temperature-dependent only above \Tep, similar to \YbCoSi.
This result implies that \YbCoRhSi ($x=0.27$) seems to be localized even though the mid-IR peak is slightly visible even at low temperature.

Thus, for Ce and Yb compounds, the behaviors of \DCOG and \DSW roughly correspond to the localized and itinerant electronic structure across the QCP.
Very close to the QCP, however, the behaviors of \DCOG and \DSW of Ce and Yb compounds are similar to each other, but they gradually change across the QCP,
suggesting the competition between the local and itinerant character.

\section{Summary}

In conclusion, we investigated the transition from the local to the itinerant states of Ce- and Yb-based heavy-fermion compounds with the same tetragonal ThCr$_2$Si$_2$-type crystal structure using the temperature dependence of the \OC spectrum.
The itinerant electronic structure appears not only in the mid-IR peak but also in the temperature dependence 
of the center of gravity of the IR \OC spectra and that of the change of the spectral weight.
The spectral-weight change also indicates that an electron-phonon interaction mainly works at high temperatures in all materials.

\section*{Acknowledgements}

We would like to acknowledge
Hiroko Yokoyama for her help in the IR experiments, 
UVSOR staff members for synchrotron radiation experiments,
Christoph Klingner for providing \YbCoSi samples,
and Christoph Geibel for his fruitful comments.
Part of this paper was supported by the Use-of-UVSOR Facility Program (BL7B) of the Institute for Molecular Science.
J.~S. and Y.~S.~K acknowledge support by the International Collaboration Program of Osaka University.


\end{document}